\documentclass[a4paper]{PoS}

\title{First study of Mrk\ 501 through the eyes of NuSTAR, VERITAS and the {\it LIDAR-corrected} eyesight of MAGIC}

\ShortTitle{Mrk\ 501 study by NuSTAR, VERITAS, and MAGIC with LIDAR correction}

\author{\speaker{Koji Noda}$^a$, Amy Furniss$^b$, Josefa Becerra Gonz\'{a}lez$^{c,d}$, Greg Madejski$^b$, and David Paneque$^a$,\\
  on behalf of {\it Fermi-LAT}, MAGIC, {\it NuSTAR}, VERITAS collaborations, and GASP-WEBT, F-GAMMA consortiums, and many campaign participants\\
 \\
  \llap{$^a$}Max-Planck-Institut f\"{u}r Physik, Germany\\
  \llap{$^b$}Stanford University, USA\\
  \llap{$^c$}NASA Goddard Space Flight Center, USA\\
  \llap{$^d$}Instituto de Astrof\'{i}sica de Canarias, Spain\\
  E-mail: \email{knoda@mppmu.mpg.de}}
\abstract{
  The blazar Mrk\ 501 is among the brightest X-ray and TeV sources in the sky, and among the few sources whose (radio to Very-High-Energy (VHE; $\geq$ 100 GeV) gamma-rays) spectral energy distributions can be characterized by current instruments by means of relatively short observations (minutes to hours). 
  In 2013, we organized an extensive multi-instrument campaign involving the participation of {\it Fermi-LAT}, MAGIC, VERITAS, F-GAMMA, {\it Swift}, GASP-WEBT, and other collaborations/groups and instruments which provided the most detailed temporal and energy coverage on Mrk\ 501 to date. 
  This observing campaign included, for the first time, observations with the Nuclear Stereoscopic Telescope Array ({\it NuSTAR}), which is a satellite mission launched in mid-2012. 
  {\it NuSTAR} provides unprecedented sensitivity in the hard X-ray range 3-79 keV, which, together with MAGIC and VERITAS observations, is crucial to probe the highest energy electrons in Mrk\ 501.
  The multi-instrument campaign covered a few day long flaring activity in July 2013 which could be studied with strictly simultaneous {\it NuSTAR} and MAGIC observations. 
  A large fraction of the MAGIC data during this flaring activity was affected by hazy atmospheric conditions, due to the presence of a sand layer from the Saharan desert. 
  These data would have been removed in any standard Cherenkov Telescope data analysis. 
  The MAGIC collaboration has developed a technique to correct for adverse atmospheric conditions the very high energy (VHE, $E>100$ GeV) observations performed by Cherenkov telescopes. 
  The technique makes use of the atmospheric information from the LIDAR facility that is operational at the MAGIC site, and applies an event-by-event correction to recover data affected by adverse weather conditions. 
  This is the first time that LIDAR information has been used to produce a physics result with Cherenkov Telescope data taken during adverse atmospheric conditions, and hence sets a precedent for current and future ground-based gamma-ray instruments.
  In this contribution we report the observational results, focusing on the LIDAR-corrected MAGIC data and the strictly simultaneous {\it NuSTAR} and MAGIC/VERITAS data, and discuss the scientific implications.
}

\FullConference{The 34th International Cosmic Ray Conference,\\
		30 July- 6 August, 2015\\
		The Hague, The Netherlands}

\begin{document}

\section{Introduction}
Mrk\ 501 ($z=0.034$) is one of the brightest blazars in the X-ray energy band, and is also known to emit very high energy (VHE; $E \geq 100$  GeV) gamma-rays. 
As well as its continuously high brightness of $>0.3$ Crab\ Nebula Unit (C.U.) in the VHE waveband, this source is known by its fast variablity, elevated even up to tens of C.U., and flickering with a time scale as short as 2 minutes\,\cite{Albert2007}. 
Due to its brightness in a wide waveband range from radio to VHE gamma rays, the spectral energy distributions (SEDs) of this source can be characterized with observations for a relatively short time scale, minutes to hours. 
Thus, recently, observational campaigns have been coordinated with participation by several instruments, with which the SED of the source can be reproduced by strictly simultaneous observations. 
Such observations are extremely important for highly variable sources such as Mrk\ 501, to reduce an ambiguity in the interpretation of SEDs observed in a short flaring state compared with an averaged SED. 

For example, a mild activity of Mrk\ 501 was observed in a campaign in 2009, which included for the first time {\it Fermi-LAT} data\,\cite{Abdo2011}, 
and the campaign revealed that the averaged SED is well described with a simple single-zone synchrotron self-Compton (SSC) model.
On the other hand, an analysis of an alternate data set, but still a part of the 2009 campaign\,\cite{Doert2013}, showed a variability up to 4.5 C.U. above 300 GeV, as observed by Whipple. This period includes at least two clear flares.
It was challenging to model one of the two flares with the standard SSC model using observations between X-ray and VHE instruments, because the correspondence was not clear between a hardening in X-ray spectrum (without catching the synchrotron peak) and varying spectra taken in VHE band. 
Multi-instrumental campaigns are still quite important to collect various multiwavelength states with simultaneous observations, to have a global picture of the source behavior. 

\section{Instruments and the observation}
In 2013 we organized an extensive multi-instrument campaign including {\it Fermi-LAT}, MAGIC, VERITAS, {\it NuSTAR}, F-GAMMA, {\it Swift}, GASP-WEBT, and other collaborations/groups. 
This provided the most detailed temporal and energy coverages on Mrk\ 501 ever, since we have started the organization of the campaigns in 2008. 
In particular, this campaign included, for the first time, observations with the Nuclear Stereoscopic Telescope Array ({\it NuSTAR}), which is a space-based hard X-ray telescope launched in 2012. 
{\it NuSTAR} provides unprecedented sensitivity in the hard X-ray range of 3-79 keV, which is of particular importance to understand properties of the highest energy electrons injected into the emission region(s) of Mrk\ 501, together with MAGIC and VERITAS observations. 
The details of the participating instuments can be found in \cite{Furniss2015}, where the analysis and an expanded interpretaton of this campaign are also described. 

\begin{figure}[h!]
  \begin{center}
    \includegraphics[width=\textwidth]{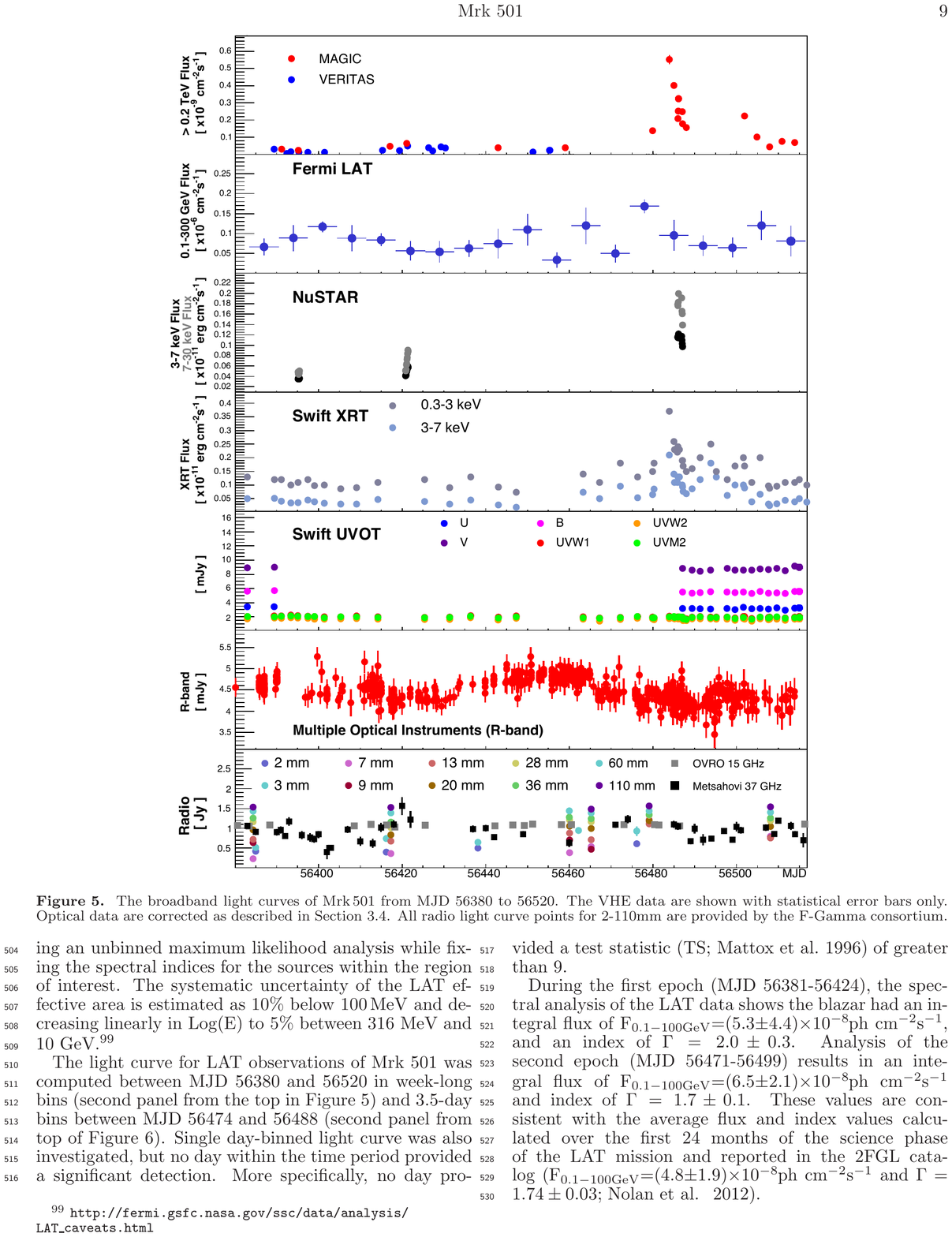}
    \caption{The broadband light curve of the Mrk\ 501 observation in 2013\,\cite{Furniss2015} (preliminary).}
    \label{fig1}
  \end{center}
\end{figure}

The campaign covered a few day long flaring activity in July 2013. In the active period, we triggered a Target-of-Opportunity (ToO) observation to get maximum overlap of strictly simultaneous observations between X-rays and VHE instruments, namely, {\it NuSTAR} and MAGIC observations. 
The overall light curve obtained is shown in Figure\,\ref{fig1}. 
A clear correlation is seen between observed flux values by X-ray satellites ({\it NuSTAR} and {\it Swift/XRT}) and by VHE instruments (MAGIC and VERITAS). 
On the other hand, {\it Fermi-LAT} observations showed a mild variability, for which it is hard to see a clear correlation with other instruments.
Possible correlations between the different frequencies were studied. A significant correlation was detected only between the X-ray and the VHE bands. 
A quadratic function is preferred with respect to a linear function, which naturally indicates that the majority of the inverse-Compton up-scattering of photons is likely to be occurring within the Thomson regime, but does not totally excludes a possibility of having it within the Klein-Nishina regime, as mentioned later. 

\section{LIDAR correction for MAGIC data}
In this campaign a large fraction of the MAGIC data were affected by a sand layer from the Saharan desert, in particular during the flaring activity (MJD 56483 and later).
Such data would have been removed in any standard Cherenkov Telescope data analysis, due to difficulties in the analysis processes. 
First, it is challenging to correct event energies of air-showers developed in the dust-ridden atmosphere, especially if the absorption happens at a height close to the shower maximum development. 
Also, it is difficult to understand the atmospheric transmission profile that is changing in a short time scale of minutes. 
The former point is not applicable for this case, as the sand layer lies only at relatively low altitude (up to 5 km, typically $<$ 3 km).
To overcome the second point, we used information from a LIDAR facility at the MAGIC site, taken during the observation once per 5 minutes. 
Then, we applied an event-by-event correction in order to reliably use these data. 
There are two steps in the correction method: one for the estimated energy of each shower event, and the other for the effective area according to the correction in the energy. 
More details of these correction processes can be found in a contribution in the previous conferences \cite{Fruck2013,Fruck2014}. 
Recently these analysis processes have been implemented in the standard analysis package in MAGIC, MARS\,\cite{Zanin2013}. 
The method was tested using Crab Nebula observations under non optimal atmospheric conditions. 

\begin{figure}[h!]
  \begin{center}
    \includegraphics[width=0.9\textwidth]{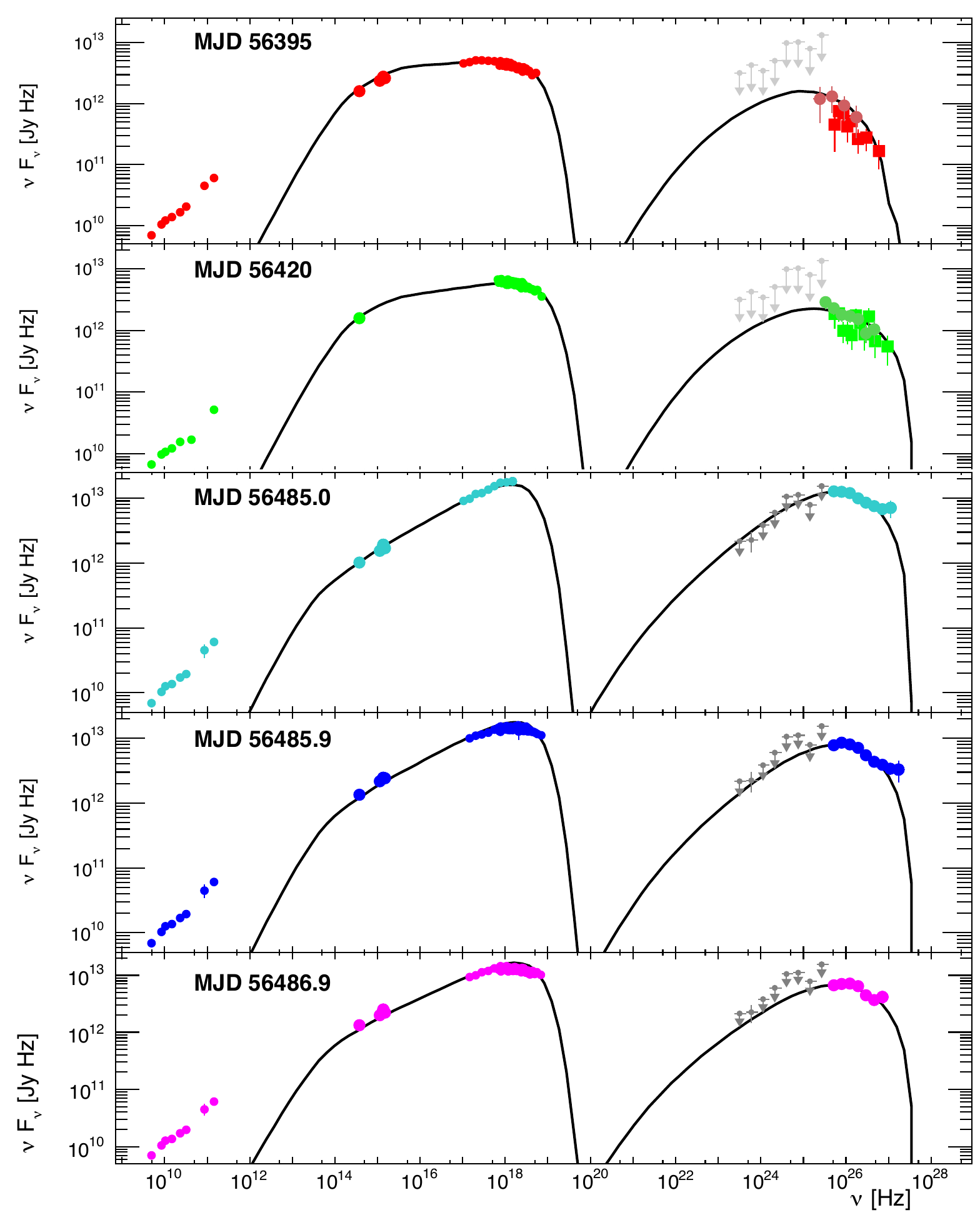}
    \caption{The broadband SED of the Mrk\ 501 observation in 2013\,\cite{Furniss2015} (preliminary).}
    \label{fig2}
  \end{center}
\end{figure}

The MAGIC data points in the overall light curve (Fig.\,\ref{fig1}) are already corrected by the LIDAR data. 
Their error bars are applied accordingly, depending on the presense/absense of the LIDAR correction in each analysis. 
The systematic error of the corrected flux values is estimated to be 15\%\footnote{The detailed discussion of the systematic error in the method can be found in \cite{Fruck2015}.}.
In the campaign we have collected about 22 hours of the MAGIC data in total, and about 17 hours out of them are affected by the sand layer. 
Thanks to LIDAR correction, more than 10 hours of the data have been recovered, and we finally obtained 15.1 hours of the usable data in total. 
The SED of the detected gama-rays is computed for each day, and is also corrected, accordingly if the day is affected by the sand layer. 
Figure\,\ref{fig2} shows the SEDs for each day, plotted together with corresponding data from all other instruments. 
The top two panels show SEDs taken in two days (MJD 56395 and 56420) in the former half of the campaign including also {\it NuSTAR} and VERITAS observations, when the source was in a relatively quiescent state. 
MAGIC and VERITAS points are colored differently in these two panels. 

The highest flux in X-ray observations was recorded in MJD 56484, and in the following day (MJD 56485) we have obtained a good dataset in X-rays ({\it Swift/XRT}) and VHE (MAGIC, with the LIDAR-correction), as shown in the middle panel. 
In the following two days we had a more complete campaign including {\it NuSTAR} observations, which are shown in the bottom two panels, where MAGIC data are again corrected by the LIDAR data. 
Note that, {\it Fermi-LAT} data (shown in grey) were averaged over a period of about a month containing the campaign, due to its limited sensitivity, and so its SEDs in the top two panels and bottom three panels are identical, respectively. 
Other observations with X-ray and VHE instruments were strictly simultaneous, which enables us to study the temporal evolution of the SEDs without any large ambiguity. 

\section{Results and conclusions}
The obtained SEDs are modelled with a simple single-zone SSC model, which is a standard in this source. 
In particular, it is modeled with an equilibrium version of the single-zone SSC model\,\cite{Boettcher2002}.
The model curves are shown by black lines in Fig.\,\ref{fig2}, and resultant parameters are seen in Tab.\,\ref{tab1}. 
Note that, radio data points are considered as upper limits in the following SED modeling, as it is well known that the source has an extended radio emission outside the region of the jet emission, but affecting the overall flux. 

The Doppler factor of the emission region can vary much from state to state. Though there are methods to overcome it, in this work we have simply fixed it to 15, chosen as in the previous studies for Mrk\ 501 (e.g., \cite{Abdo2011}). 
Also, another parameter, $\eta$, that determines the escape time scale of the injected particles (by $t_{\rm esc} = \eta R/c$) was fixed to 100, motivated by a success in previous studies for TeV blazars (e.g., \cite{Aliu2013}). 
As a result, the temporal evolution of the fitting parameters showed a hardening in the injected particle from the relatively quiescent state on MJD 56420 ($q=1.8$) to an elevated state on MJD 56485 ($q=1.3$). 
Accordingly, the magnetic field ($B$) of the emission region decreased from 0.05 to 0.03 G, and the equipartition parameter (a ratio of the Poynting flux carried by the magnetic field to the electron kinetic energy, related to the equilibrium particle distribution but not the injected) decreased down to 0.001, which is far from the equipartition. 
All the parameter behaviours are consistent with a picture that the energy in the magnetic field transferred to the acclerated electrons, to be dominant. 
One note should be added that the hardening in the injected electrons is difficult to be reconstructed in the standard shock acceleration mechanism, suggesting a possible application of other models such as a magnetic reconnection event (e.g., \cite{Romanova1992}). 
A small decrease of the emission region size ($R$, from 7.0 to 5.0, in the unit of $10^{15}$cm) seen in the model fitting is also consistent with this picture. 
The above modeling indicates that the inverse-Compton scattering of the photons near the synchrotron peak is far into the Klein-Nishina regime. 
This is not necessarily in opposition to the indication by the quadratic correlation found between X-ray and VHE instruments, as the quadratic correlation can occur even in the Klein-Nishina regime. 
Such interpretations of the model fitting were discussed in the conference, and are discussed also in \,\cite{Furniss2015}. 
\begin{table}
  \begin{tabular}{c|ccccc}
    \hline
    \hline
    Parameter & MJD 56395 & MJD 56420 & MJD 56485.0 & MJD 56485.9 & MJD 56486.9\\
    \hline
    $\gamma_{min}\,[\times 10^4]$ & 1.5 & 2.1 & 2.0 & 2.0 & 2.0\\
    $\gamma_{max}\,[\times 10^6]$ & 1.0 & 1.4 & 1.4 & 1.7 & 1.4\\
    $q$       & 1.9 & 1.8 & 1.3 & 1.3 & 1.3\\
    $\eta$    & 100 & 100 & 100 & 100 & 100\\
    $B\,[{\rm G}]$ & 0.06& 0.05& 0.03& 0.03& 0.03\\
    $\Gamma$  & 15  & 15  & 15  & 15  & 15\\
    $R \, [\times 10^{15} {\rm cm} ]$ & 7.0 & 7.0 & 5.0 & 7.0 & 7.0\\
    $\theta \, [{\rm degrees}]$ & 3.8 & 3.8 & 3.8 & 3.8 & 3.8\\
    $L_e \, [{\rm erg/cm^2/s}]$ & $9\times 10^{42}$ & $12\times 10^{42}$ & $36\times 10^{42}$ & $28\times 10^{42}$ & $26\times 10^{42}$\\
    $\epsilon = L_B / L_e$ & $1.8 \times 10^{-2}$ & $6.1 \times 10^{-2}$ & $5.3 \times 10^{-4}$ & $1.3 \times 10^{-3}$ & $1.4 \times 10^{-3}$\\
  \end{tabular}
  \caption{Single-zone SSC model parameter values.}
  \label{tab1}
\end{table}

This is the first time that LIDAR information is used to produce a physics result with Cherenkov telescope data taken during adverse atmospheric conditions. 
The result shows the data can be well corrected and used without any special treatment other than a reasonable increase to the systematic error. 
This work sets a precedent for the current and future ground-based gamma-ray instruments, such as Cherenkov Telescope Array.

\acknowledgments{
We would like to thank all the supports to {\it NuSTAR}, MAGIC, VERITAS, {\it Fermi-LAT}, {\it Swift}, and all the other optical and radio instruments. The full acknowledgments can be found in the paper\,\cite{Furniss2015}.

This work was supported under NASA Contract No.\ NNG08FD60C, and made use of data from the NuSTAR mission, a project led by the California Institute of Technology, managed by the Jet Propulsion Laboratory, and funded by the National Aeronautics and Space Administration. 
We thank the NuSTAR Operations, Software and Calibration teams for support with the execution and analysis of these observations. 
This research has made use of the NuSTAR Data Analysis Software (NuSTARDAS) jointly developed by the ASI Science Data Center (ASDC, Italy) and the California Institute of Technology (USA).

MAGIC would like to thank the Instituto de Astrof\'{\i}sica de Canarias for the excellent working conditions at the Observatorio del Roque de los Muchachos in La Palma. 
The financial support of the German BMBF and MPG, the Italian INFN and INAF, the Swiss National Fund SNF, the ERDF under the Spanish MINECO (FPA2012-39502), and the Japanese JSPS and MEXT is gratefully acknowledged. 
This work was also supported by the Centro de Excelencia Severo Ochoa SEV-2012-0234, CPAN CSD2007-00042, and MultiDark CSD2009-00064 projects of the Spanish Consolider-Ingenio 2010 programme, by grant 268740 of the Academy of Finland, by the Croatian Science Foundation (HrZZ) Project 09/176 and the University of Rijeka Project 13.12.1.3.02, by the DFG Collaborative Research Centers SFB823/C4 and SFB876/C3, and by the Polish MNiSzW grant 745/N-HESS-MAGIC/2010/0.

VERITAS is supported by grants from the U.S. Department of Energy Office of Science, the U.S. National Science Foundation and the Smithsonian Institution, and by NSERC in Canada. We acknowledge the excellent work of the technical support staff at the Fred Lawrence Whipple Observatory and at the collaborating institutions in the construction and operation of the instrument. 
The VERITAS Collaboration is grateful to Trevor Weekes for his seminal contributions and leadership in the field of VHE gamma-ray astrophysics, which made this study possible.

The {\it Fermi-LAT} Collaboration acknowledges generous ongoing support from a number of agencies and institutes that have supported both the development and the operation of the LAT as well as scientific data analysis. 
These include the National Aeronautics and Space Administration and the Department of Energy in the United States, the Commissariat \`{a} l'Energie Atomique and the Centre National de la Recherche Scientifique / Institut National de Physique Nucl\'{e}aire et de Physique des Particules in France, the Agenzia Spaziale Italiana and the Istituto Nazionale di Fisica Nucleare in Italy, the Ministry of Education, Culture, Sports, Science and Technology (MEXT), High Energy Accelerator Research Organization (KEK) and Japan Aerospace Exploration Agency (JAXA) in Japan, and the K. A. Wallenberg Foundation, the Swedish Research Council and the Swedish National Space Board in Sweden. 

Additional support for science analysis during the operations phase is gratefully acknowledged from the Istituto Nazionale di Astrofisica in Italy and the Centre National d'\'{E}tudes Spatiales in France. 

We thank the {\it Swift} team duty scientists and science planners and we acknowledge the use of public data from the Swift data archive. This research has made use of the {\it Swift} XRT Data Analysis Software (XRTDAS) developed under the responsibility of the ASI Science Data Center (ASDC), Italy. 

}


\begin{thebibliography}{99}

\bibitem{Albert2007} 
J. Albert et al., {\it Astrophysical Journal}, {\bf 669}, 862, 2007. 
\bibitem{Abdo2011} 
A. Abdo et al., {\it Astrophysical Journal}, {\bf 727}, 129, 2011. 
\bibitem{Doert2013}
M. Doert et al., {\it Proc. of 33rd ICRC (Rio de Janeiro)}, \#0762, 2013.  
\bibitem{Furniss2015} 
A. Furniss. K. Noda, et al., {\it Astrophysical Journal}, submitted, 2015. 
\bibitem{Fruck2013} 
C. Fruck et al., {\it Proc. of 33rd ICRC (Rio de Janeiro)}, \#1054, 2013. 
\bibitem{Fruck2014} 
C. Fruck and M. Gaug, {\it Proc. of ATMOHEAD 2014 (Padova)}, 02003, 2015. 
\bibitem{Zanin2013} 
R. Zanin et al., {\it Proc. of 33rd ICRC (Rio de Janeiro)}, \#0773, 2013. 
\bibitem{Fruck2015} 
C. Fruck, Ph.D. Thesis (Technische Universit\"{a}t M\"{u}nchen), 2015. 
\bibitem{Boettcher2002} 
M. B\"{o}ttcher and J. Chiang, {\it Astrophysical Journal}, {\bf 581}, 127, 2002. 
\bibitem{Aliu2013} 
E. Aliu et al., {\it Astrophysical Journal}, {\bf 779}, 92, 2013. 
\bibitem{Romanova1992} 
M. Romanova and R. Lovelace, {\it A\&A}, {\bf 262}, 26, 1992. 


\end{thebibliography}
\end{document}